\documentclass{emulateapj}

\newcommand{\hi}{\mbox{\rm \ion{H}{1}}}
\newcommand{\hii}{\mbox{\rm \ion{H}{2}}}
\newcommand{\htwo}{\mbox{\rm H$_2$}}

\newcommand{\xco}{\mbox{$X_{\rm CO}$}}
\newcommand{\xcounits}{\mbox{cm$^{-2}$ (K km s$^{-1}$)$^{-1}$}}

\newcommand{\sighi}{\mbox{$\Sigma_{\rm HI}$}}
\newcommand{\sightwo}{\mbox{$\Sigma_{\rm H2}$}}
\newcommand{\siggas}{\mbox{$\Sigma_{\rm gas}$}}

\newcommand{\rtf}{\mbox{\rm r$_{25}$}}

\shorttitle{A Universal Neutral Gas Profile for Nearby Disk Galaxies}
\shortauthors{Bigiel \& Blitz}
\begin{document}

\slugcomment{Accepted for publication in the Astrophysical Journal}
\title{A Universal Neutral Gas Profile for Nearby Disk Galaxies} 

\author{F.~Bigiel\altaffilmark{1} and L.~Blitz\altaffilmark{2}}

\altaffiltext{1}{Institut f\"ur theoretische Astrophysik, 
Zentrum f\"ur Astronomie der Universit\"at Heidelberg,
Albert-Ueberle Str. 2, 69120 Heidelberg, Germany; bigiel@uni-heidelberg.de}
\altaffiltext{2}{Department of Astronomy, Radio Astronomy Laboratory,
University of California at Berkeley, CA 94720, USA}

\begin{abstract}

Based on sensitive CO measurements from HERACLES and \hi\ data from THINGS, we
show that the azimuthally averaged radial distribution of the neutral gas surface
density (\sighi + \sightwo) in 33 nearby spiral galaxies exhibits a well-constrained 
universal exponential distribution beyond $0.2\times\rtf$ (inside of which the scatter is large) with less than a factor of two scatter out to 
two optical radii \rtf. Scaling the radius to \rtf\ and the total gas surface density to
the surface density at the transition radius, i.e., where \sighi\ and \sightwo\ are
equal, as well as removing galaxies that are interacting with their environment,
yields a tightly constrained exponential fit with average scale length 0.61$\pm$0.06\,\rtf.
In this case, the scatter reduces to less than 40\% across the optical disks
(and remains below a factor of two at larger radii). We show that the tight
exponential distribution of neutral gas implies that the total neutral gas mass of nearby disk
galaxies depends primarily on the size of the {\it stellar} disk
(influenced to some degree by the great variability of
\sightwo\ inside $0.2\times\rtf$). The derived prescription predicts the total gas mass
in our sub-sample of 17 non-interacting disk galaxies to within a factor of two. Given the
short timescale over which star formation depletes the \htwo\ content of these
galaxies and the large range of \rtf\ in our sample, there appears to
be some mechanism leading to these largely self-similar radial gas distributions
in nearby disk galaxies.
\end{abstract}


\section{Introduction} \label{intro} Scaling relations are important
because they are often thought to be manifestations of some underlying
fundamental physical processes. They also make it possible to
determine properties of a system with incomplete measurements to
within the scatter of the scaling.  
The total neutral gas content of galaxies has long defied such
characterizations because of the wide diversity of previously observed
gas surface density distributions \citep[see, e.g.,][]{helfer03}. This was at least partly
due to the lack of comprehensive, resolved, extended and deep
observations of atomic (\hi) and molecular gas (\htwo) across the
disks of a significant sample of galaxies. This limitation has been
overcome by the availability of new radio surveys providing sensitive
\hi\ and CO (the standard tracer for \htwo) radio data, tracing the
gas content of many nearby galaxies out to large galactocentric radii 
\citep[THINGS, HERACLES,][]{walter08,leroy09}.

With these data sets in hand, we can now re-assess the azimuthally
averaged gas distributions (radial profiles) in nearby spiral
galaxies. Both HI and CO profiles tend to show quite distinct
behavior: the HI surface density distributions are often depressed in
the centers, increasing in the inner disk and slowly declining in
the outer regions.  Furthermore, the surface density in the outer
regions can be affected by environmental factors such as interaction
with companions or truncation due to the ram pressure stripping by the
hot gas in galaxy clusters \citep{cayatte94}. As for the molecular
gas, it tends to fall off much more sharply than the HI in an
approximately exponential fashion that varies from galaxy to galaxy
\citep{young82,regan01,leroy08,bigiel08}.

We show in this paper, however, that when the totality of the neutral
(i.e., atomic and molecular) gas in normal spirals is considered based on the data
from these new surveys, and the profiles are properly scaled, the gas exhibits a radial profile
shape that is remarkably constant from galaxy to galaxy. The gas surface density is found to vary by no more than
about a factor of two from this average profile across the optical disk and even beyond. We
show that the scatter reduces even further when we eliminate galaxies that are
interacting with their environment.

\section{Data \& Methodology} 
\label{data}

In this paper we use radial profiles (i.e., azimuthal averages in
tilted rings) of atomic hydrogen, \sighi, molecular hydrogen (as
traced by CO emission), \sightwo, and the total gas surface density
\siggas=\sighi+\sightwo. We derive the \htwo\ profiles out to one and the
\hi\ and total gas profiles from the centers out to
two optical radii \rtf, which is defined as the 25\,mag\,arcsec$^{-2}$
B-band isophote. We will refer to the regime within \rtf\ as the optical disk.

The VLA THINGS survey \citep[The HI Nearby Galaxy Survey,][]{walter08},
provides sensitive ($\sigma_{\rm HI} \approx
0.5$~M$_\odot$~pc$^{-2}$ at 30$\arcsec$ resolution),
extended ($\sim0.5\degr$ field-of-view) HI
data. These data, together with new and archival data for some of the
target galaxies, allow us to track the HI distribution out to
$2\times\rtf$ with good sensitivity.

The IRAM 30m survey HERACLES provides the distribution of $^{12}$CO(2-1)
emission in our sample galaxies \citep[HERA CO Line Extragalactic Survey,][]{leroy09}.
HERACLES provides sensitive ($\sigma_{\rm H2}
\approx5$~M$_\odot$~pc$^{-2}$ for the most distant, face-on systems)
data covering the entire optical disk out to \rtf. The quoted \htwo\
surface densities in this paper include a Galactic CO-to-\htwo\
conversion factor $\xco=2\times10^{20} \xcounits$ and a CO line ratio
$I_{CO}(2-1)/I_{CO}(1-0)=0.7$ \citep[see][]{leroy12}. Because \hi\ can be observed out to
much larger radii for individual lines-of-sight, \citet{schruba11}
used the velocity information from the \hi\ as a prior to stack
individual CO spectra. This technique allows one to probe down to
$\sigma_{\rm H2} \approx1$~M$_\odot$~pc$^{-2}$ and thus to derive
radial CO profiles out to the edge of the optical disk ($\sim\rtf$).
This represents a significant improvement over previous studies and we
adopt their radial \htwo\ profiles derived from this approach.

We thus focus our analysis on the data and galaxy sample presented in
\citet{schruba11}, which consists of 33 nearby, star-forming disk
galaxies (excluding edge-on, low-mass and low-metallicity systems).
All quoted surface densities in this paper include the contribution from helium (a
factor of 1.36) and are corrected for inclination.  We refer the
reader to \citet{schruba11} for further details and references
regarding sample selection and properties, \hi\ and CO datasets, the
stacking technique for the CO data and the conversion of line
intensities to surface densities.

We stress that it is the combination of such deep, wide-field CO data,
the stacking technique as well as the sensitive HI data, which permits
to derive unprecedentedly accurate radial distributions of the atomic,
molecular and total gas across galaxy disks. These radial profiles are
shown for each galaxy in the galaxy atlas in the Appendix of
\citet{schruba11} (note, however, that we extend their \hi\ profiles
out to 2$\times$\rtf).

We also include radial gas profiles for the Milky Way for comparison.
We adopt the \hi\ and \htwo\ profiles from \citet{dame93}\footnote{We use their \hi\ profile based on
the \citet[][KBH]{kulkarni82} rotation curve and the
\htwo\ profile following \citet{digel91} at large radii.}, adjust the
latter to match our adopted value for \xco\ and include the
contribution from helium.  We assume $\rtf=16$\ kpc,
based on the observation that \hii\ regions, planetary nebulae and
carbon stars seem to disappear at about 2$\times$\,R$_\sun$ in the
Milky Way \citep{fich84, schneider83, schechter88}.

\section{Results}
\label{results}
\subsection{Unscaled Radial Profiles}
\label{unscaled}

\begin{figure*}
\epsscale{0.72}
\plotone{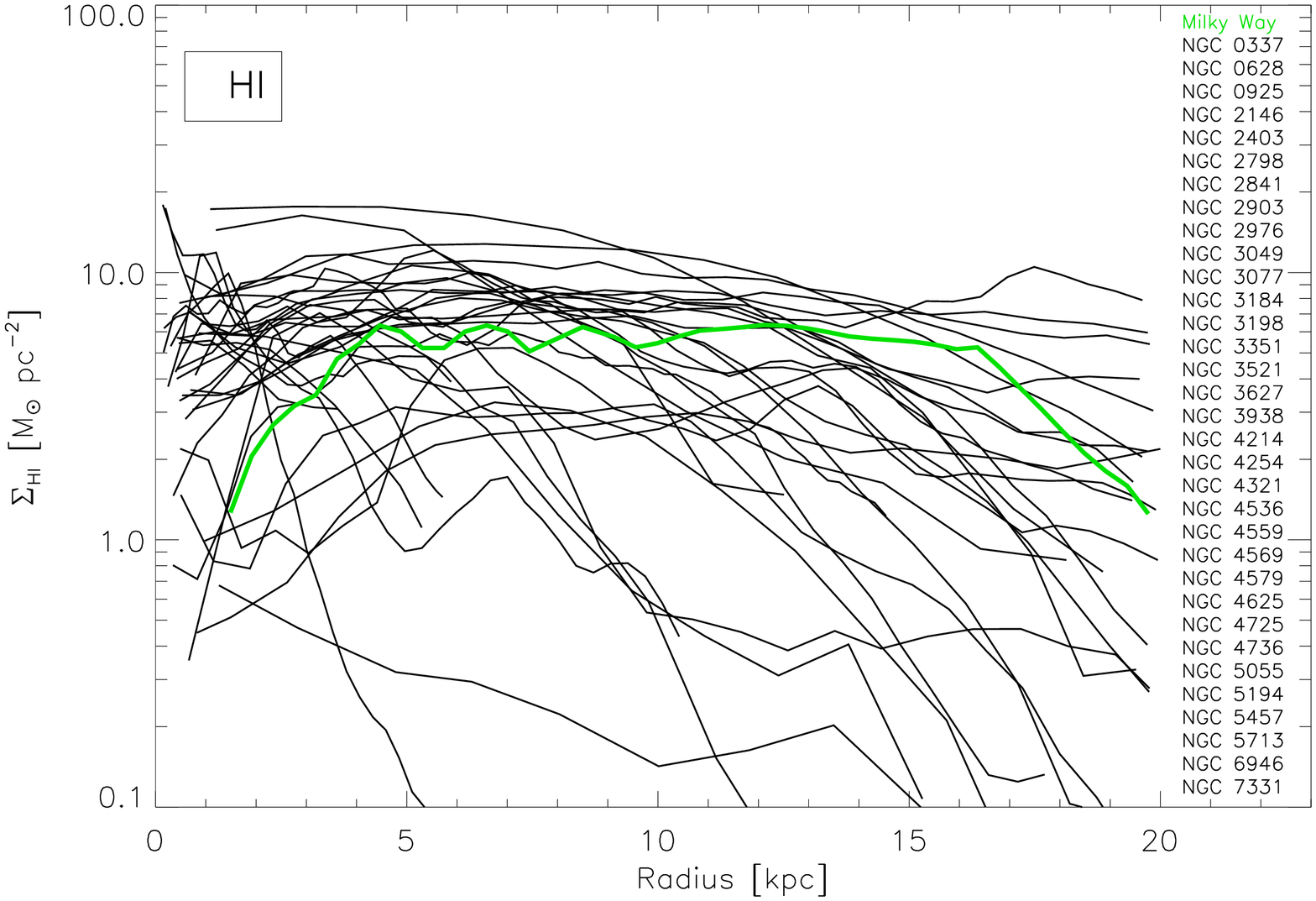}
\plotone{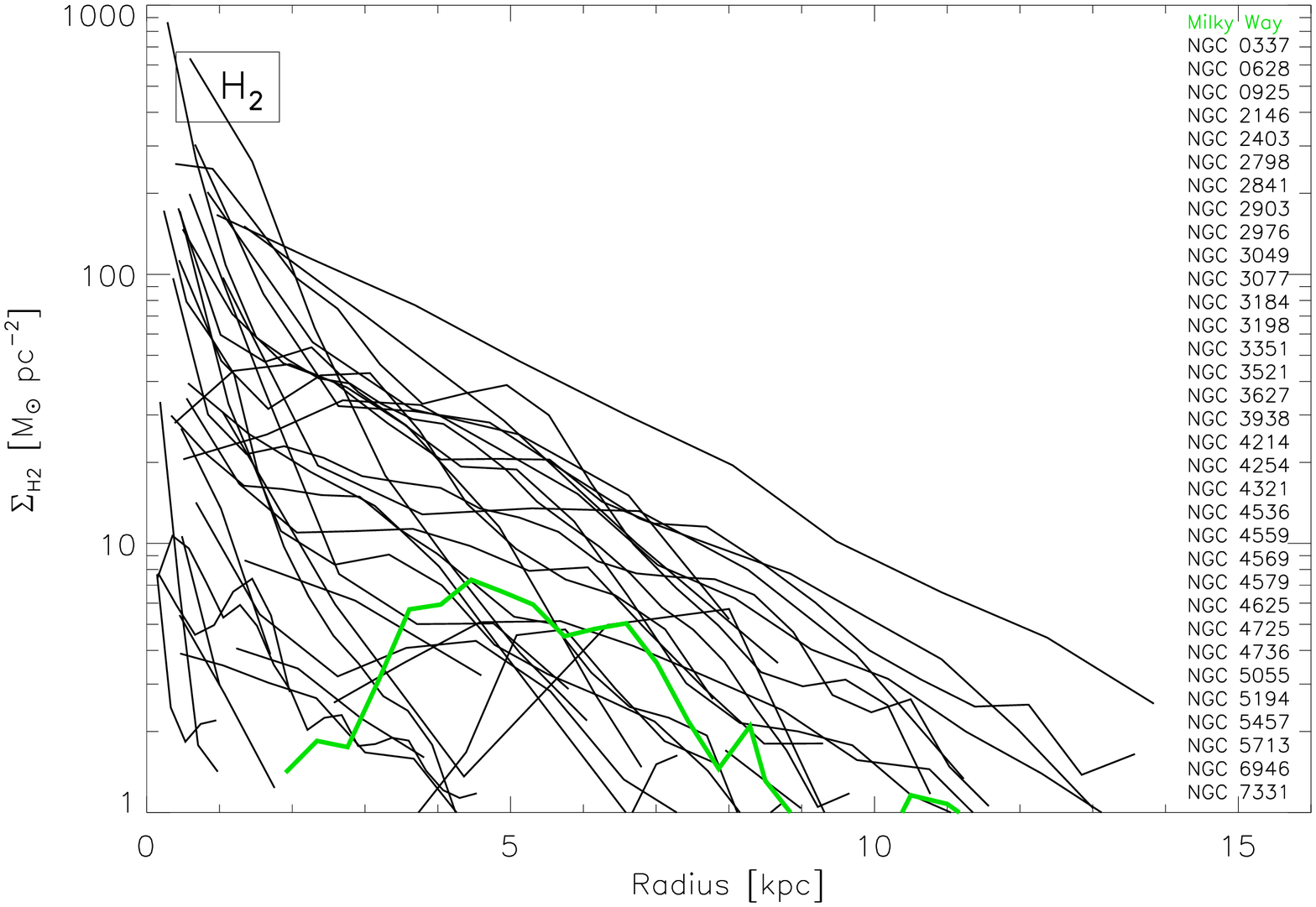}
\plotone{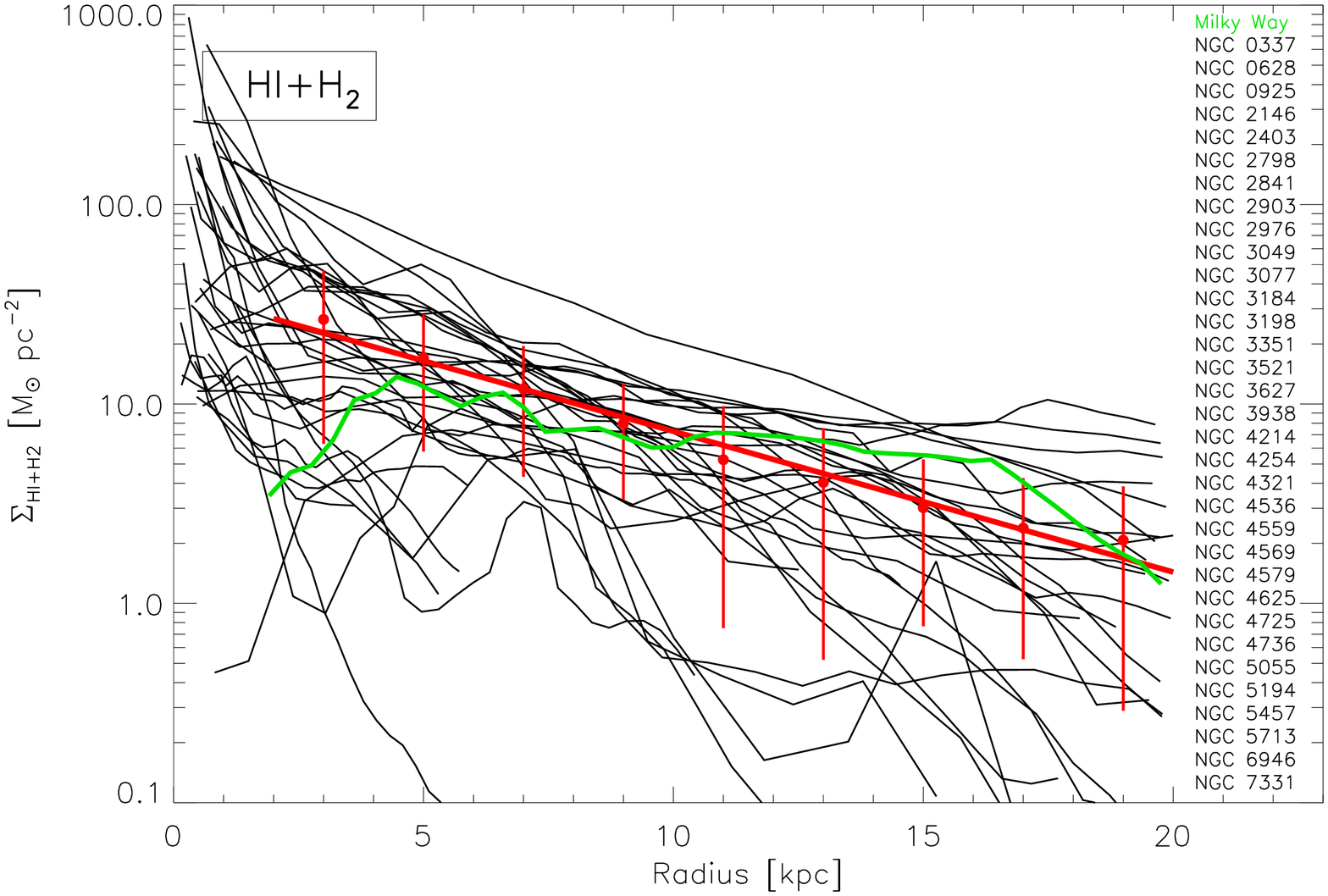}
\caption{Radial profiles of the surface densities of atomic gas (\sighi, top panel), molecular gas
(\sightwo, middle panel) and total gas (\siggas=\sighi+\sightwo, bottom panel) for 33 nearby,
star-forming disk galaxies and the Milky Way. While the \hi\ profiles are generally flat and concentrate just below
10 M$_\odot$~pc$^{-2}$, the \htwo\ profiles decline approximately exponentially. An exponential fit to the binned means
(red points; error bars show the RMS scatter) characterizes the bulk of the total gas profiles well and the scatter about the fit remains
within a factor of two even at large radii.}
\label{fig1}
\end{figure*}

Figure \ref{fig1} shows the \hi, \htwo\ and total gas surface density
profiles of the 33 sample galaxies and the Milky Way. 
Radially slightly less extended versions of these profiles as well as the \hi\
and CO maps they are derived from are shown for each of our sample
galaxies in \citet{schruba11}. We refer the reader to the
Appendix of that paper to match individual galaxies to particular
profiles.
  
The top panel of Figure \ref{fig1} shows the \hi\ profiles. They are generally
relatively flat and show a great deal of scatter with little apparent
regularity except for a concentration just below $\sim$10\,M$_\odot$~pc$^{-2}$
for radii $r\lesssim10-12$\,kpc. This is the maximum \hi\ surface density (for
near solar-metallicity galaxies) before the interstellar medium (ISM) becomes primarily
molecular \citep{martin01,wong02,leroy08,bigiel08}. Note that the
profiles fill almost all of the ``phase space'' between 1 and 10 
M$_\odot$~pc$^{-2}$ at all radii.  Some profiles are
truncated and decrease steeply: these are often interacting systems or
cluster members.  In the inner parts, many galaxies show a more or
less pronounced depression. These central ``\hi\ holes'' are usually
regions where the ISM is predominantly molecular \citep[the
\sightwo/\sighi\ ratio increases roughly exponentially with decreasing
radius; e.g.,][]{leroy08,bigiel08}. The Milky Way \hi\ profile agrees well with
the overall trend: \hi-deficient in the center and relatively flat
throughout at $\sighi\approx5-7$\ M$_\odot$~pc$^{-2}$.

Almost all \htwo\ profiles in the middle panel show an
exponential-like
decline. The scale length, however, is different from
galaxy-to-galaxy, i.e. some profiles are steeper than others. Like the
\hi\ profiles, the \htwo\ profiles fill almost all of the ``phase space'' up
to the most molecule-rich galaxy. In their inner parts, many galaxies
deviate from this exponential trend and the \htwo\ surface density
rises much more steeply \citep[see also][]{regan01}. Even when
averaged over many square kpc, some galaxies reach surface densities
of many 100\,M$_\odot$~pc$^{-2}$ in the inner parts, quite similar to what is
observed in starburst galaxies \citep[e.g.,][]{kennicutt98}. The Milky
Way profile also declines roughly exponentially from $r\approx5$\,kpc on
outward, but shows a depression at smaller radii. Where it declines
exponentially, it does so with a scale length typical of many of the
other galaxies.

The plot of the total gas profiles is given in the bottom panel and
shows that the combination of the relatively flat \hi\ profiles (over
large parts of the radius range considered here) and the exponential
\htwo\ profiles leads to somewhat more coherent total gas profiles for
many of the galaxies than do either the \hi\ or \htwo\ profiles alone.
The total gas profiles cluster at surface densities between a few and a
few times 10 M$_\odot$~pc$^{-2}$ between $r\approx2-12$\,kpc.
The Milky Way profile falls quite in the middle
of the bulk of the profiles and thus its averaged gas distribution
matches that of most of the other galaxies well.

We also show mean data values (red points) and associated 1$\sigma$
RMS scatter of the profiles in 2\,kpc wide bins (note that the means
are plotted in the middle of these bins). We exclude the central 2\,kpc
for the binning where the profiles rise much more steeply than further out.
The relative scatter ranges
from $\sim60\%$ to $\sim80\%$ within the optical disks, i.e., for
$r<\rtf$, and remains within a factor of two even at large radii where
the profiles begin to flare significantly.

The red line shows an ordinary-least-squares (OLS) fit to the means
and yields an average scale length of 6.1\,kpc with standard deviation
0.3\,kpc. Even though the fit is relatively well constrained
(reflected by the small standard deviation), we estimate the impact of
several other factors on the fit result by doing the following: 1)
we add noise to each data point (mean) and re-fit, 2) we bootstrap
(sampling profiles from the ensemble allowing repetition), 3) we vary
the radius range to fit over and 4) we vary the bin size we use to
derive the means. The first method is an uncertainty estimate based on
the error on the mean in each bin, the second addresses how much
individual profiles drive the fit and the last two approaches probe
the impact of our specific choices for radius range and bin size.
Adding noise as well as varying radius range and bin size lead to an
uncertainty $<$0.5\,kpc. Bootstrapping leads to an uncertainty of
$\sim0.7$\,kpc, which we adopt as our uncertainty on the fit so that
we quote the mean scale length of the unscaled total gas profiles as
$6.1\pm0.7$\,kpc. We will show in the following that the agreement
among these profiles can be improved further by scaling both axes the
right way and by focusing on non-interacting galaxies.

\subsection{Scaled Radial Profiles}

By measuring the radius in physical units (kpc), intrinsic differences
in the size of the galaxies will lead to scatter in the profile
distribution. This can be accounted for if the x-axis (radius) is
normalized by the optical radius (\rtf) of the galaxy. This scaling
reduces the scatter to $40-60\%$ across the optical disk (beyond which
it reaches up to 100\% at large radii, similar to the unscaled
profiles). This is notably less than the $60-80\%$ in the unscaled
case above.

\citet{blitz04} noted, that there is a natural scale to
the gas in normal spirals if the conversion from \hi\ to \htwo\ is
governed primarily by the midplane hydrostatic pressure in galactic
disks.  In that case, the location where the gas in the disk goes from
being primarily molecular in the inner regions, to where it becomes
primarily atomic in the outer regions, should occur at a constant {\it
stellar} surface density. They showed that for the 28 galaxies they
analyzed, this constancy is good to about 40\% and has a value of $\sim$120
M$_\sun$ pc$^{-2}$. \citet{leroy08} reached the same conclusion
(although they derived a stellar surface density of
$81\pm25$ M$_\sun$ pc$^{-2}$), using a combination of HERACLES, THINGS and BIMA SONG
\citep[BIMA Survey of Nearby Galaxies,][]{helfer03} data for 23 nearby
galaxies. We determine this transition radius (i.e., where $\sighi = \sightwo$)
directly for each galaxy from
the radial profiles (typically the transition occurs at $\siggas\approx14$\,M$_\sun$ pc$^{-2}$).

\begin{figure*}[!t]
\epsscale{0.7}
\plotone{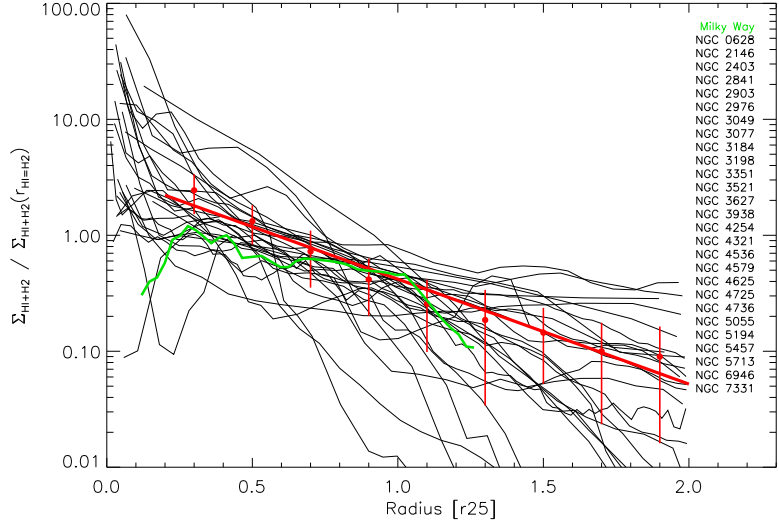}
\caption{Scaled radial profiles for the galaxy ensemble: the radius (x-axis) is normalized to the
size of the optical disk and \siggas\ (y-axis) is scaled to the value at the transition radius, i.e., where the
ISM transitions from being primarily molecular to being primarily atomic. The combined distribution
of profiles is remarkably tight, including the Milky Way profile which is right
in the middle of the galaxy ensemble. The exponential fit to the means is well constrained and the scatter
is reduced significantly compared to the unscaled profiles in the bottom panel of Figure \ref{fig1}.}
\label{fig2}
\end{figure*}

In Figure \ref{fig2} we show the profiles one obtains
if both scalings are applied: the radius is normalized to the size of the optical disk (x-axis)
and the gas surface density is scaled to the value of \siggas\ at the transition radius (y-axis). 
For six galaxies the transition radius cannot be readily determined: the ISM in these galaxies is
either entirely dominated by \hi\ (NGC\,337, NGC\,925, NGC\,4214, NGC\,4559) or
entirely \htwo\ dominated (NGC\,4569, NGC\,2798). We will not include these galaxies 
in this plot and the following.

\begin{figure*}[!t]
\epsscale{0.7}
\plotone{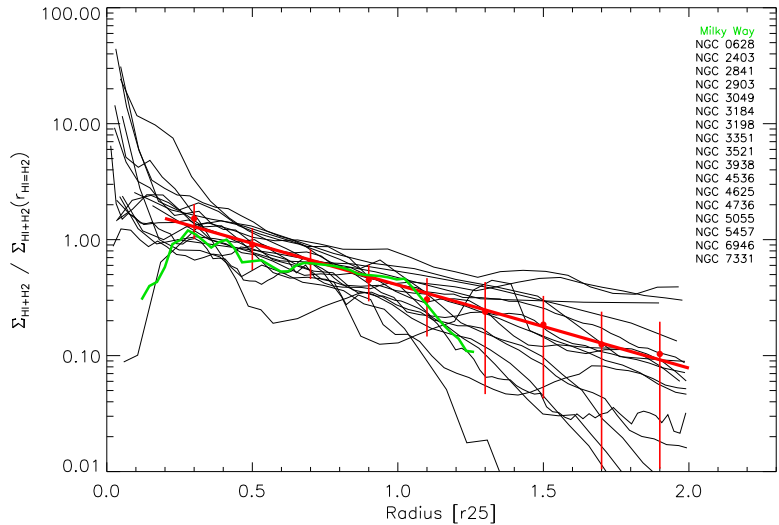}
\caption{Scaled radial profiles as in Figure \ref{fig2}, but after removing galaxies showing signs of interaction with their
environment. The distribution of these disk galaxy profiles is extremely tight, with a scatter across the optical disk of only $25-40\%$ and
a scale length of the exponential fit of $\sim0.61\,\rtf$. The Milky Way profile is an excellent match to the average trend.}
\label{fig3}
\end{figure*}

The red points in Figure \ref{fig2} show mean values and 1$\sigma$ scatter in 0.2$\times$\rtf-wide
bins. The scatter increases from $\sim35\%$ to $\sim80\%$ at large
radii where the profiles begin to flare. Across the optical disks the
scatter is between $35\%$ and $50\%$.  While the normalization of the x-axis
has a bigger impact reducing the scatter, the additional scaling of the surface density
reduces the scatter further to $50\%$ or less within the optical disks and to
less than a factor of two even at large radii. The minimum scatter of
$35\%$ is almost a factor of two improvement compared to the minimum
scatter of $60\%$ for the unscaled case.  This matches the visual
impression that the scaled profiles in Figure \ref{fig2} show a
remarkably tight distribution.

The red line shows the OLS fit to the means and yields an average
scale length of $0.48\pm0.04$\rtf, where the error is the standard
deviation. Following the approach described in Section \ref{unscaled}, we estimate the uncertainty
of the fitted scale length in several ways. We find that bootstrapping
and varying the radius range dominate the uncertainties and we quote our
final fit result as $0.48\pm0.06$ \rtf.

The gas distribution, particularly in the outer parts, can be
significantly affected by interactions with other galaxies in groups
or clusters or by ram pressure stripping from a hot intracluster
medium. Oftentimes these interactions result in truncated \hi\
profiles, driving the scatter in the profile ensemble. 

In addition to the six galaxies with uncertain transition radii above,
we further eliminate 10 galaxies that exhibit evidence for
interactions with their environments. These are galaxies that show
signs of strong tidal interactions in their HI distributions and those
which are members of the Virgo or Coma clusters subject to
ram-pressure stripping. These galaxies are: NGC\,2146, NGC\,2976, NGC\,3077, NGC\,3627, NGC\,4254,
NGC\,4321, NGC\,4579, NGC\,4725, NGC\,5194, NGC\,5713 (note that
NGC\,2798 and NGC\,4569 would be part of this list too, but are
already excluded based on the unclear transition radius in these
galaxies).

Figure \ref{fig3} shows the plot that results if both axes are scaled
as in Figure \ref{fig2}, but after removing the interacting galaxies.
For the remaining 17 galaxies, the scatter within the optical disks reduces to $25-40\%$ and
remains between $\sim80-90\%$ beyond \rtf. The Milky Way profile
is a good match to the overall trend. The ensemble constitutes a
tight, well-defined exponential distribution across the optical disk with
deviations in the centers and flaring in the outer disks. The fit
yields a scale length of $0.61\pm0.03$ \rtf\ and we quote our final
result based on bootstrapping as the dominant source of uncertainty as
$0.61\pm0.06$ \rtf. 

\section{Discussion}
\label{discussion}

Figure \ref{fig3} shows that galaxies not interacting with their
environment exhibit a tight exponential distribution of their total
gas content when the radius is scaled to the optical radius and the
surface density of the gas is scaled to the surface density of the gas
at the transition radius. The scaling relation for disk galaxies we
derive is:

\begin{equation}
\frac{\siggas}{\Sigma_{\rm trans}} = 2.1\times e^{-1.65\times r/r_{25}},
\label{hih2}
\end{equation}

\noindent
where {\mbox{$\Sigma_{\rm trans}$}} is the surface density of the gas
at the transition radius. The factors $2.1$ and $-1.65$ are obtained
from the y-intercept and slope of the exponential fit in Figure 3.
This implies that for these disk galaxies the total mass of
neutral gas, ${\rm M_{gas}}$, is given by

\begin{equation}
{\rm M_{gas}} = 2\pi\times 2.1\times \Sigma_{\rm trans} \times \rtf^2 \times {\rm X},
\label{mgas}
\end{equation}

\noindent
where the factor X depends on how far out from the center one integrates Equation
\ref{hih2}. Integrating out to the optical radius \rtf\ yields ${\rm X} = 0.18$,
out to 2$\times$\rtf\ yields ${\rm X} = 0.31$ and integrating to infinity yields
${\rm X} = 0.37$.  

Equation \ref{mgas} shows that for any given maximum radius, ${\rm M_{gas}}$
depends only on ${\rtf}^2$ and ${\Sigma_{\rm trans}}$. ${\Sigma_{\rm trans}}$ does not vary a great deal from galaxy to galaxy
and has a typical value of about 14\,M$_\odot$~pc$^{-2}$ \citep[also compare][]{leroy08,bigiel08}.
Thus, ${\rm M_{gas}}$ depends primarily on ${\rtf}^2$, which varies by two orders of
magnitude for the galaxies plotted.

This leads to the surprising result that the mass of neutral
hydrogen gas depends mostly on the size of the {\it stellar} disk and that
the gas arranges itself somehow into a distribution that is self-similar among
galaxies. The result also implies that except for the region
of a galaxy at $\lesssim0.2\,\rtf$, which is highly variable, the total
neutral gas mass of a disk galaxy at $z = 0$ can be estimated if \rtf, i.e., the extent of
the stellar disk, is known.  The
variable inner regions of galaxies, are, however, quite small, and
despite either depressions or large excesses of the molecular gas in these
regions, they typically only account for a small fraction of the total
neutral gas mass ($\sim15\%$ on average in our sample).

\begin{figure*}[!t]
\epsscale{0.6}
\plotone{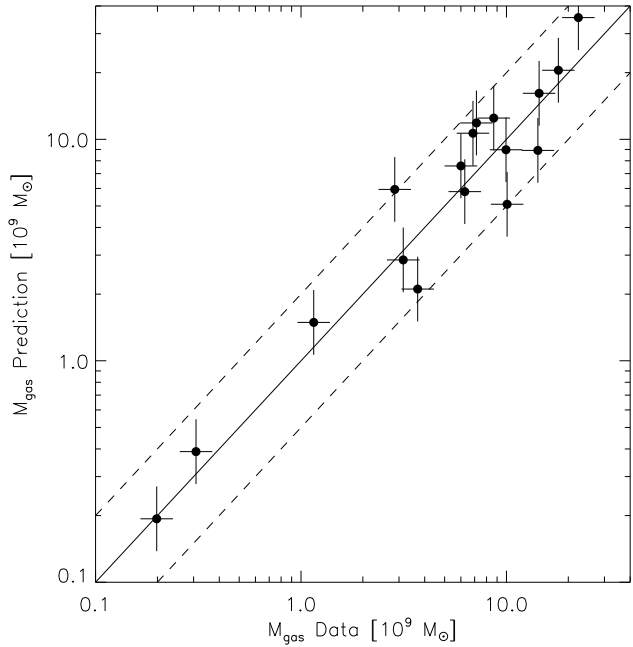}
\caption{Total (\hi+\htwo) gas mass obtained from the average radial profile fit (Figure \ref{fig3}
and Equation \ref{mgas}) versus the mass obtained directly from the \hi\ and \htwo\
intensity maps. The flux is computed out to two optical radii \rtf\ for each of the non-interacting galaxies
shown in Figure \ref{fig3}. The solid line indicates unity and the two dashed lines a 
factor of two scatter. The plot thus shows that Equation \ref{mgas} allows us to predict the total gas mass of a disk galaxy within a factor of two uncertainty.}
\vspace{0.59cm}
\label{fig4}
\end{figure*}

We examine this conclusion in Figure \ref{fig4}, which shows
the total gas mass of the galaxies in Figure \ref{fig3} as measured in the
respective \hi\ and \htwo\ intensity maps, compared to the predictions
made using Equation \ref{mgas} (where the integration is carried out to 2$\times$\rtf).
We estimate the uncertainty on the total gas masses derived from the intensity maps (x-axis)
from the calibration uncertainties of the THINGS and HERACLES data
\citep[$5\%$ and $20\%$, respectively,][]{walter08, leroy09}. We thus adopt $20\%$ as the
overall uncertainty. For the masses
derived from the fit (y-axis), we estimate the uncertainty from error propagation (from Equation \ref{mgas}).
We take into account the uncertainties on the fit (slope and intercept), 
\rtf\ \citep[estimated from][]{paturel91} and 
${\Sigma_{\rm trans}}$ \citep[from][]{leroy08}. This results in an uncertainty estimate of $\sim40\%$ for
the mass prediction. Both uncertainties are indicated as error bars in Figure \ref{fig4}.

Figure \ref{fig4} shows that this equation offers a robust way to
predict the total gas mass of disk galaxies to within a factor of two (indicated by the
dashed lines, the solid line indicates unity). The mean ratio of predicted versus measured
total gas mass is $1.17$ (with standard deviation $0.36$), which underlines the good
correspondence between prediction and data. This close correspondence is 
rooted in the tightly
constrained fit in Figure \ref{fig3}, which falls almost directly
on the means in the individual radius bins.

Based on observed metallicity gradients across galaxy disks 
\citep[e.g.,][]{moustakas10}, one might speculate about a similarly radially
varying CO-to-\htwo\ conversion factor. Recent work indeed suggests that
systematic variations of \xco\ with radius (albeit not necessarily driven
primarily by metallicity) occur in at least some of the
galaxies studied in this paper: in these cases, \xco\ is observed to
increase with increasing radius (Sandstrom et al., in prep.). A low
value for XCO in galaxy centers seems to be a more common feature
(Sandstrom et al., in prep.). The former effect would lead to
shallower \htwo\ profiles , while the latter might in fact explain the
observed apparent \htwo\ excesses in galaxy centers, possibly even giving
rise to truly exponential \htwo\ profiles including the centers. 

One of the implications of a universal gas profile is that there
appears to be some mechanism that keeps the relationship constant in
the face of star formation in these galaxies.  For example,
\citet[]{bigiel08,leroy08, bigiel11} show that the depletion time for the
molecular gas in disk galaxies is $\sim 2 \times 10^9$\,yr. The region of
Figure 3 where $\frac{\Sigma_{\rm gas}}{\Sigma_{\rm trans}}>1$ 
(roughly for $r< 0.45\times\rtf$) is
the region dominated by molecular gas and that gas will therefore be
exhausted in about 15\% of a Hubble time. In order to keep the profiles
self-similar, this implies that either new gas
comes from outside the disk and falls preferentially in the central
regions, an unlikely occurrance, or that gas flows through the disk to make up the gas lost to
star formation, which takes place preferentially within $r< 0.45\times\rtf$.
In either case, it is difficult to understand why the total gas
profiles would be so self-similar.  For the infall case, one expects
such infall to be sporadic, possibly occurring in the form of small galaxies merging with the
bigger disk galaxies in our sample. For the inflow case, Figure \ref{fig3} encompasses a
variety of galaxy types and morphologies, and it is hard to see how inflow could be so
closely regulated to produce the tight observed relationship.

As cosmological simulations of galaxy evolution including gas
become ever more refined, it will become necessary for these
simulations to reproduce the universal relationship when carried out
to the present epoch.  Because of the limitations in making
observations of the atomic gas to higher redshift with present day
instrumentation, it will be difficult in the near future to extend the
work presented here to normal disk galaxies at significantly higher redshifts.  It will also be
a challenge to extend this work to lower metallicity systems, where
CO emission becomes an increasingly poor tracer of the molecular hydrogen
\citep[e.g.,][]{bolatto11}. 
On the other hand, \rtf\ can be determined in many galaxies to higher redshifts and
Equation 2 could be used to estimate M$_{\rm gas}$. 
As it becomes possible to measure {\it total} neutral gas   
masses to higher redshifts with ALMA, the JVLA, and Arecibo, direct comparisons of
the predicted and measured total mass of neutral gas can be extended both to higher
redshift and to a larger sample of local galaxies providing good tests
of the universality of the neutral gas profile presented in this
paper. 

The universal gas profile we have described seems to be a fundamental
property of normal disk galaxies at $z=0$. We have, however, only probed
galaxies of near solar metallicity and none of the galaxies in our
final sample are dwarfs. Also, \citet{young11} and
\citet{serra12} have shown recently that a surprisingly large fraction of 
early type galaxies, i.e., ellipticals and lenticulars,  
contain large amounts of atomic and molecular gas.  Whether these
galaxies obey the same universal gas profile is still to be
determined.

\acknowledgements 
We thank T.~Dame for providing a copy of his Milky Way data compilation and
A.~Schruba for making available his CO stacking data. This work was supported 
by Sonderforschungsbereich SFB 881 ``The Milky Way System''
of the German Research Foundation (DFG) and NSF grant No. 1140031 to the University of 
California at Berkeley.

\end{document}